# A study of the effects of Gaussian distribution of coherence length of source on the diffraction of partial temporal coherence beam from multi slits: Theory and simulation

**E. KOUSHKI, S.A. ALAVI***

*Department of Physics, Hakim Sabzevari University (HSU), Sabzevar, 96179-76487, Iran*

*Corresponding author: s.alavi@hsu.ac.ir**

**Abstract:**

First, we have generalized the notion of Franhoufer diffraction of temporal coherent light from a single slit to the case of arbitrary *n*-slits. The diffraction pattern is investigated for different values of recently [19] introduced parameter $\eta$ "decoherence parameter". It is shown that for multi-slits, the temporal decoherence effects appear for $\eta \geq 1$ i.e., the coherence length is shorter that the size of the slits. Results of our study reproduce the previous studies for perfect temporal case, when coherence length tends to infinity. Then to bring the problem closer to reality, we do not fix the coherence length and consider a Gaussian distribution function for it. Numerical study of the effects of coherence parameters of Gaussian distribution on far field diffraction pattern is performed.

## 1. Introduction

Diffractive multi-slits and Optical gratings are periodic systems which have very important applications in science and technology, such as spectrometers, spectrophotometers, pulse compression gratings and spectral domain optical coherence tomography (SD-OCT) [1-3]. Their potential in separating different frequencies of light marks them as the most useful devices in light spectroscopy [4,5]. They are well-known instruments in optical spectroscopy of materials and light sources in both laboratory instruments and telescopes [6,7]. Light diffraction from double and multi slits has been the subject of many studies because due to its importance in the study of light properties, slits dimensions and optical measurements [8-11]. The technology of production and improvement of optical gratings is being extended using modern manufacturing potential of planar technologies such as lithography [12,13]. But the role of the light source coherence length has not been properly studied in the literature. The effects of temporal and spatial coherence on the diffraction from a few slits have been the subject of some researches [14-16]. Recently, the technology of measuring the temporal coherence properties of different sources has been extended. The electromagnetic degree of temporal coherence and the coherence time for quasi-monochromatic unpolarized light beams emitted by some different sources are studied in [17]. Moreover, the propagation of partially polarized and partially coherent beams in uniaxial crystals have been investigated in [18]. Recently, we studied Fraunhofer diffraction from a single slit [19]. A general formula was derived to describe the intensity distribution of the Fraunhofer diffraction pattern of a slit aperture illuminated with partial temporal coherent light and the effects of coherence length on the diffraction pattern was studied using the introduced parameter $\eta$ (decoherence parameter).

Moreover, the model was generalized to the case of circular aperture and the effects of the coherence length on the diffraction pattern were investigated. In addition, the case that the coherence length is not constant but has a Gaussian distribution has been recently investigated for the single slit and the circular aperture [20].

In this work, we extend our studies (theory and simulation) to the case of multi slits and investigate the effects of Gaussian distribution of the coherence length on the diffraction pattern resulting from multi slits. One of the reasons for considering Gaussian distribution for coherence time (and correspondingly for coherence length) in gases, is thermal collisions. In thermal sources, each atom or molecule emitting energy collides with other atoms along its path, which causes random phase jumps. The time between two consecutive collisions can be statistically represented by the Gaussian function. Between two successive collisions, the beam coming out from the source can be considered coherent, and therefore the output wave trains also have a Gaussian coherence time distribution. Therefore, it seems necessary to study the sources with Gaussian coherence time distributions.

## 2. Theory

In our previous work, the diffraction of a partial temporal coherent beam from a single-slit was studied. A general formula was derived for the intensity distribution from a single slit illuminated with partial temporal coherent light, at the far-field [19]. Theory was based on interference of waves from a source traveling along different paths. According to the fact that partial temporal coherence of light influences the interference intensity and considering a single slit as a continuous series of spot sources, we obtained an analytical relation for the intensity distribution and then generalized the model to the circular apertures [19].

In this paper we generalize our previous studies to the case of multi slits. We consider geometrically a grating as a combination of several slits. Then we not only improve the theory of the previous paper, but also extend it to the case of multiple slits. The more real case that the coherence length has Gaussian distribution is also studied and the effects of this distribution on the diffraction pattern is investigated using Monte Carlo method.

The superposition of $N$ beams with the same intensity $I$ and angular frequency $\omega$ at the far field point $P$ is given by [19]:

$$I_P = NI_s + 2I_c \sum_{j=1}^{N-1} (N-j)(1-\frac{j|\tau|}{\tau_0})\cos(j\omega\tau) \tag{1}$$

where $\tau$ and $\tau_0$ are the constant time difference (time dilation) between two neighboring paths, and the coherence time, respectively. $I_s$ is the self-product intensity of each individual segment and $I_c$ represents the cross-product intensity due to the contribution of two mutual intensities. To apply this relation to an open slit, we consider the single slit as a continuous series of spot light sources. Using calculations discussed in [19] and Appendix A and by including the constant factor $\frac{c\varepsilon_0}{2}$, the total intensity distribution from a single-slit is as follows:

$$I_P(\theta) = \frac{c\varepsilon_0}{2}[\frac{b^2}{3}(\frac{E_L}{r})^2 + 2b(\frac{E_L}{r})^2 \int_{y=0}^{y=b} (1-\frac{y}{b})(1-\frac{y|\sin\theta|}{b}\eta)\cos(ky\sin\theta)dy] \tag{2}$$

where $b$ and $E_L$ are the slit width and the amplitude per unit width of slit at unit distance correspondingly. The dimensionless parameter $\eta$ which we name it "slit decoherence parameter" is defined as $\eta = \frac{b}{l_0}$ and $l_0 = c\tau_0$, $k = \omega/c$ are the length of finite wave train and the wave number respectively. Note that in the regime of Franhoufer diffraction theory, $\sin\theta$ is enough smaller than one so that the second parentheses in the integral is always positive.

By introducing:

$$\alpha = \frac{1}{b}, \beta = \frac{|\sin\theta|}{l_0}, \gamma = k\sin\theta \tag{3}$$

the analytical solution of the integral in Eq. (2) is provided in Appendix B, so we arrive at the following expression for the intensity:

$$I_P(\theta) = \frac{c\varepsilon_0}{2}[\frac{b^2}{3}(\frac{E_L}{r})^2 + 2b(\frac{E_L}{r})^2(\frac{1}{\gamma} - \frac{2\alpha\beta}{\gamma^3})\sin(\gamma y) - (\frac{\alpha+\beta}{\gamma})y\sin(\gamma y) - (\frac{\alpha+\beta}{\gamma^2})\cos(\gamma y) +$$

$$(\frac{\alpha\beta}{\gamma})y^2\sin(\gamma y) + (\frac{2\alpha\beta}{\gamma^2})y\cos(\gamma y)] \tag{4}$$

This is an exact expression for intensity distribution, which enable us to find exact results for the intensity of multi slits with given geometries.

Now, we generalize Eq. (2) to the case of $n$-slits which $n$ may be even or odd. To do this we take the irradiating width of the slits as $w = na - t$, where $a$ is the slits separation and $t$ is the dark part between two neighboring slits ( $t = a - b$ ). For even $n$ we place the origin of the coordinates as shown in Fig. (1). The integral should be taken over all the width of slits and the upper and lower bounds of the integral are $w/2$ and $-w/2$, respectively. For even number of slits, as shown in Fig.1, the lower bound of each slit is given by:

$$y_j^{(lower)} = \begin{cases} (2j-1)\frac{t}{2} + (j-1)b & , j > 0 \\ (2j+1)\frac{t}{2} + jb & , j < 0 \end{cases} \tag{5}$$

where $j$ is an integer and labels the consecutive slits ($-n/2 \le j \le n/2, j \ne 0$), and $t$ is the dark part between two neighboring slits ( $t = a - b$ ).

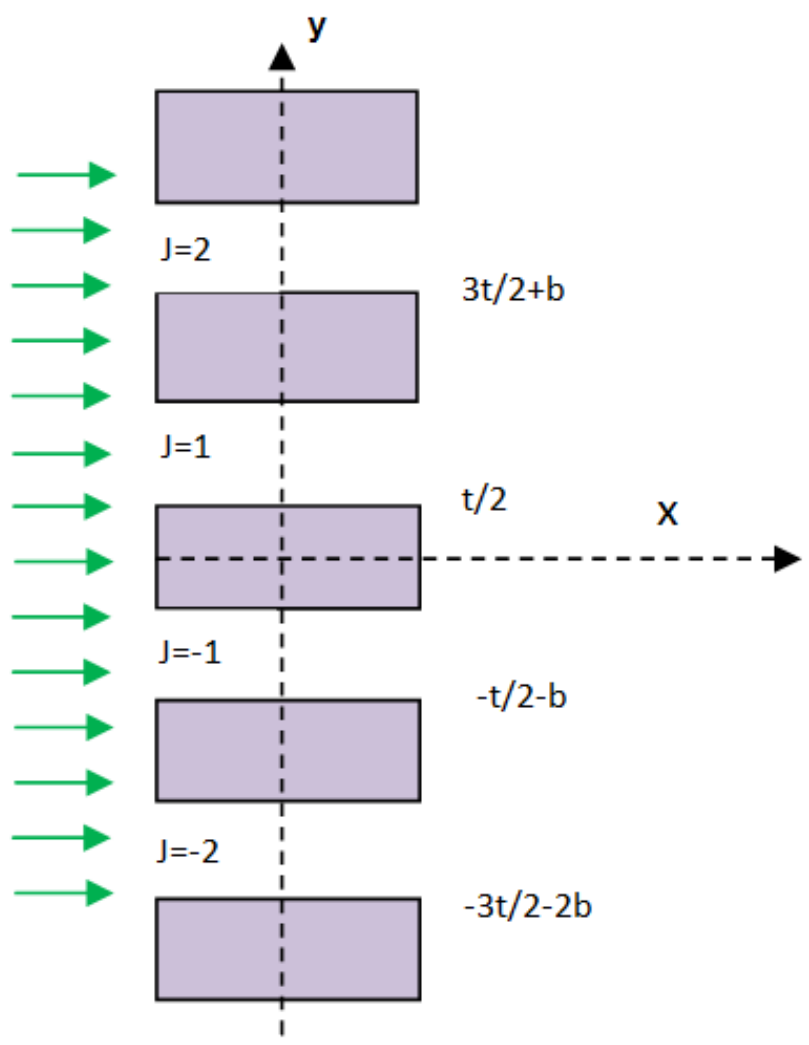

Fig.1. Geometry of multi slits diffraction with even number of slits.

For odd number of slits, the lower bound of each slit is as follows (Fig.2):

$$y_j^{(lower)} = (2j-1)\frac{b}{2} + jt \quad , {}^{-(n-1)}\!\big/_{2} \leq j \leq {}^{(n-1)}\!\big/_{2} \tag{6}$$

In both geometries the upper bound of each slit is given by $y_j^{(upper)} = y_j^{(lower)} + b$.

It is worth mentioning that, for the case of a single slit, in Eq.(2), $\alpha = 1/b$ and for multi slits it is taken as $\alpha = 2/w$.

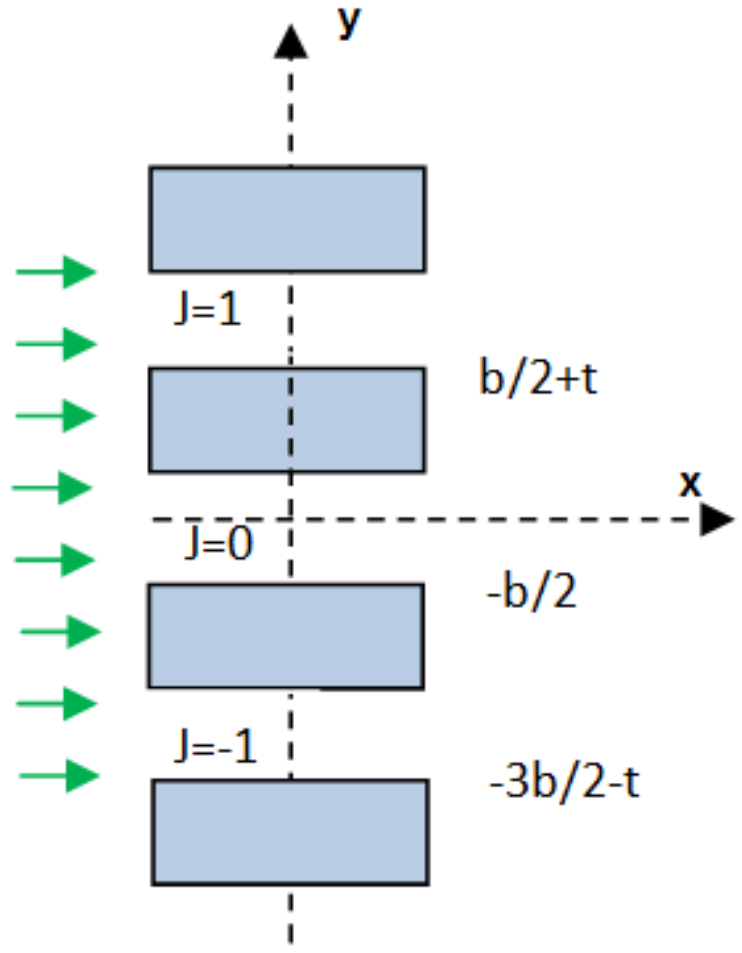

Fig.2.Geometry of multi slits diffraction with odd number of slits.

The last theoretical consideration is $r$ i.e., the distance from the slits to a typical point on the screen as shown in Fig (3), which is given by $r = d / \cos\theta$.

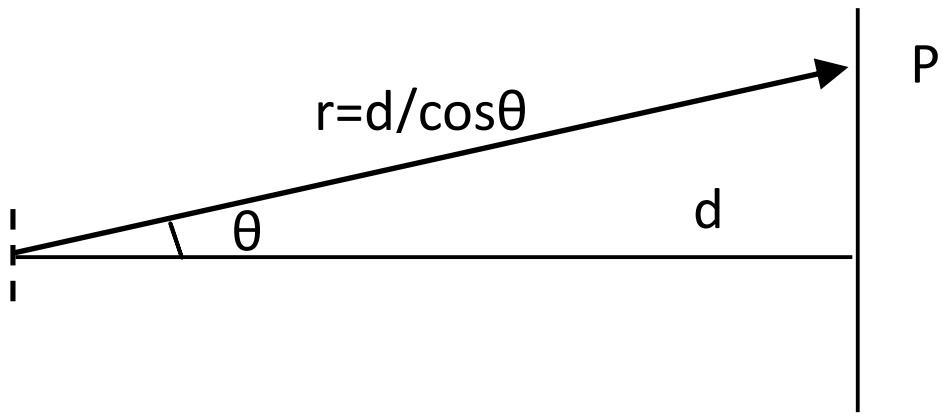

Fig.3. The distance of point P on the screen from the slits is $r = d / \cos\theta$.

The theoretical model discussed so far contains the effects of partial temporal coherence effects on the diffraction distribution pattern and could employ to study the intensity distribution at far field.

## 3. Results and discussion:

First, we check our general expression (4), for the case of a single slit $n$=1, by comparing it with our previous work [19]. Results are shown in Fig.(4). The intensities are plotted in terms of variable $\beta = \frac{kb\sin(\theta)}{2}$. As it is observed, the matching is excellent over the whole regions. In the case of perfect coherence ($\eta \to 0$), the model recovers the well-known fact that the first minimum is occurred at $\beta_0 \approx \pi rad$ (see e.g., [11]).

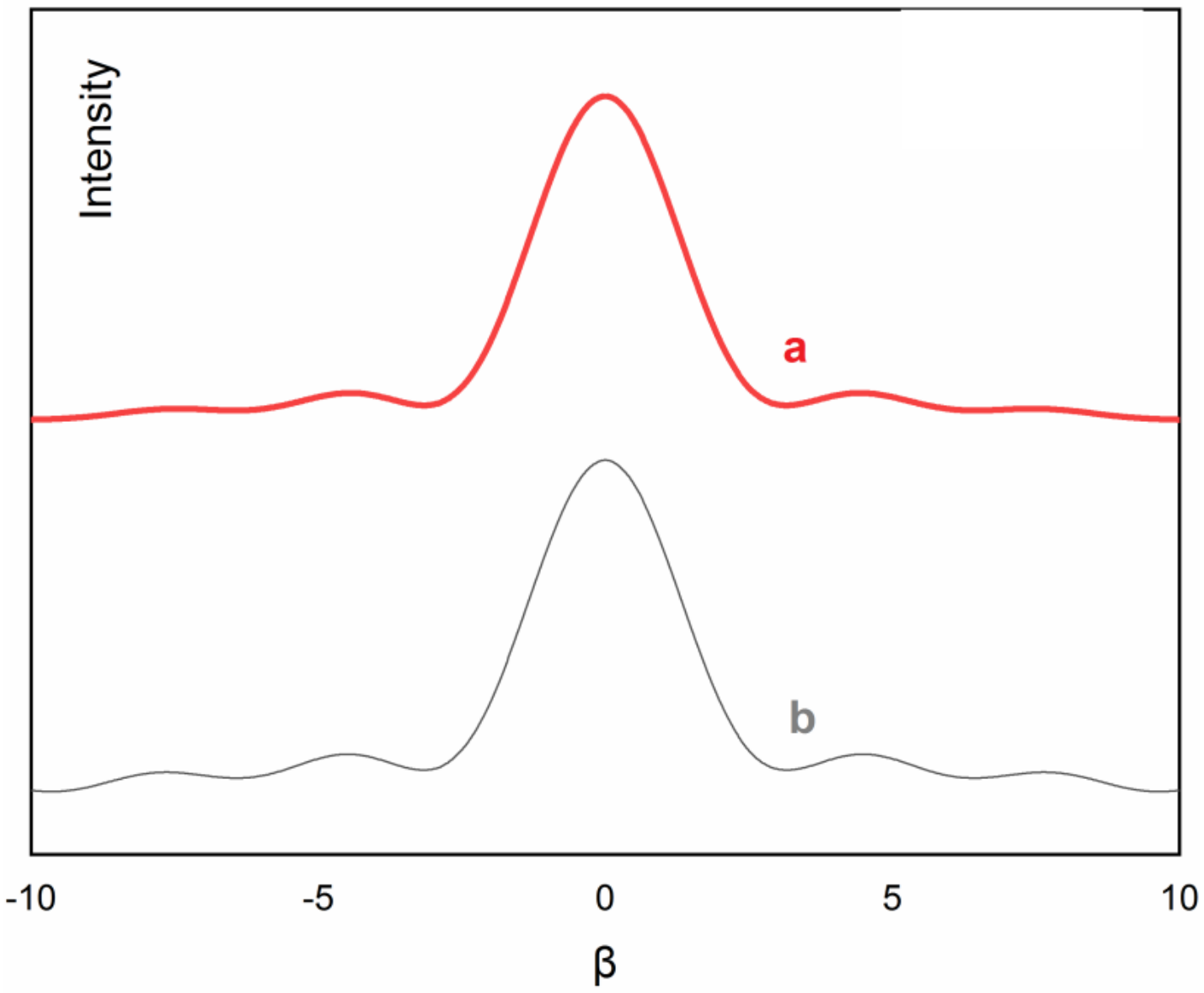

Fig.4. Comparison of Eq. (4) for n=1 with the results of our previous work [19]. (a) and (b) represent the results of our current and previous works respectively. The slit with b=50 micron is illuminated by a He-Ne laser beam with wavelength of 633 nm at far-field *r*=250cm.

The case of perfect coherence is compared with partial coherence in Fig. (5). As shown, for short coherence length ( $\eta \to \infty$ ) the first order diffraction peaks gradually disappear (Fig. (5).b).

Now, we consider the case that $l_0$ is not constant and obeys a Gaussian distribution (we name it now *l*, with mode of Gaussian distribution $l_0$ ). It is well-known that, programming languages can generate uniform random numbers in the interval 0 to 1. To generate random values for Gaussian distribution, the following relation is used [21,22]:

$$y = \sqrt{-2Ln(x_1)} + \cos(2\pi x_2) \tag{7}$$

where $x_1$ and $x_2$ are two uniform distribution between 0 and 1 and y has a Gaussian distribution with the full width at half maximum (FWHM) about $w_0 = 2.3$ and centered at 0 (the mode value) [20]. In Fig.5, Gaussian distributions of variable y have been plotted for different numbers of n (the number of recalling Eq. (7) in Monte Carlo method).

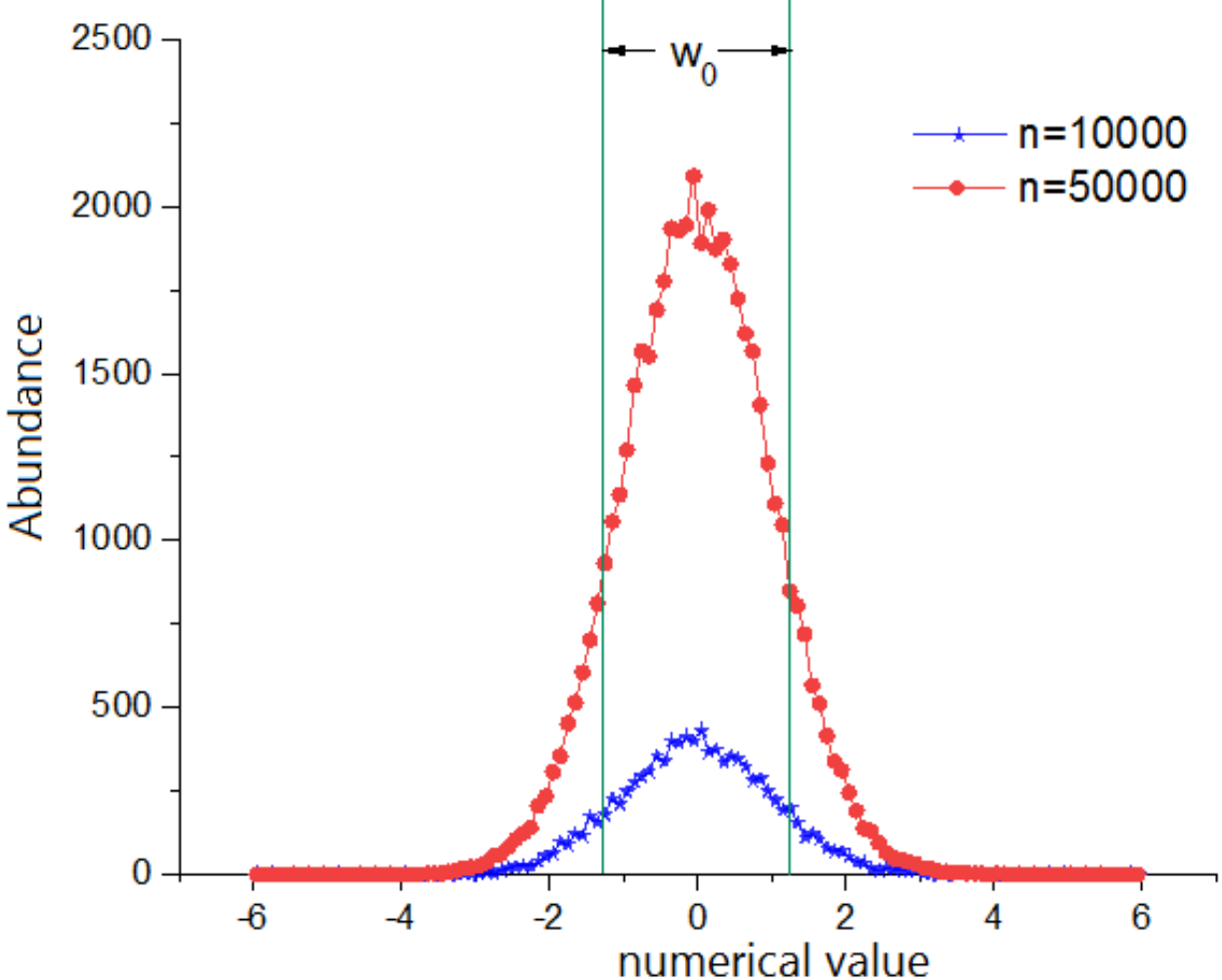

Fig.5. Gaussian distributions of “y” obtained using Eq. (7), for different “n”.

To generate a Gaussian distribution for coherence length $l$ with mode of $l_0$ and FWHM of $w$, we use the following relation:

$$l = l_0(1+\Delta.y) \quad (8)$$

where $\Delta$ is a coefficient, the bigger " $\Delta$ " is, the wider the coherence length distribution will become. Also, $y$ is the above mentioned Gaussian distribution i.e., (Eq.7) and $l$ has a Gaussian distribution with FWHM of $w = l_0 w_0 \Delta = 2.3 l_0 \Delta$

.

In the next step, the diffraction pattern of partial coherent beams with Gaussian distribution of coherence length will be obtained using equations (4) and (8). To do this, we put $l$ values extracted from Eq. (8) in Eq. (4) and get the resulting Fraunhofer diffraction distribution for a small fraction of a second. It is carried out for several hundred to several thousand times, and each $l$ (belonging to the Gaussian distribution) contributes in the diffraction pattern for a specific fraction of time. It is worth mentioning that the diffraction pattern that we observe is the temporal average of the instant patterns, which, of course, is completely stable and contains the effects of the Gaussian distribution of the coherence [20].

In Fig.(6), diffraction pattern of a single slit is plotted for three cases. The first curve (a) shows the perfect coherence, which is also shown in Fig.(4). In curve (b), coherence length is constant ($\Delta = 0$) and shorter than the slit width ($\eta = 10$) and its destructive role is appeared in the first order peak. In (c) and (d) cases, coherence length has a Gaussian distribution ($\Delta \neq 0$). It is observed that, by increasing the width of distribution (curve d), the destructive roll is more visible in the first peak. It should be noted that destructive effects of partial coherence beam on the interference patterns have been observed by other researchers [23-24].

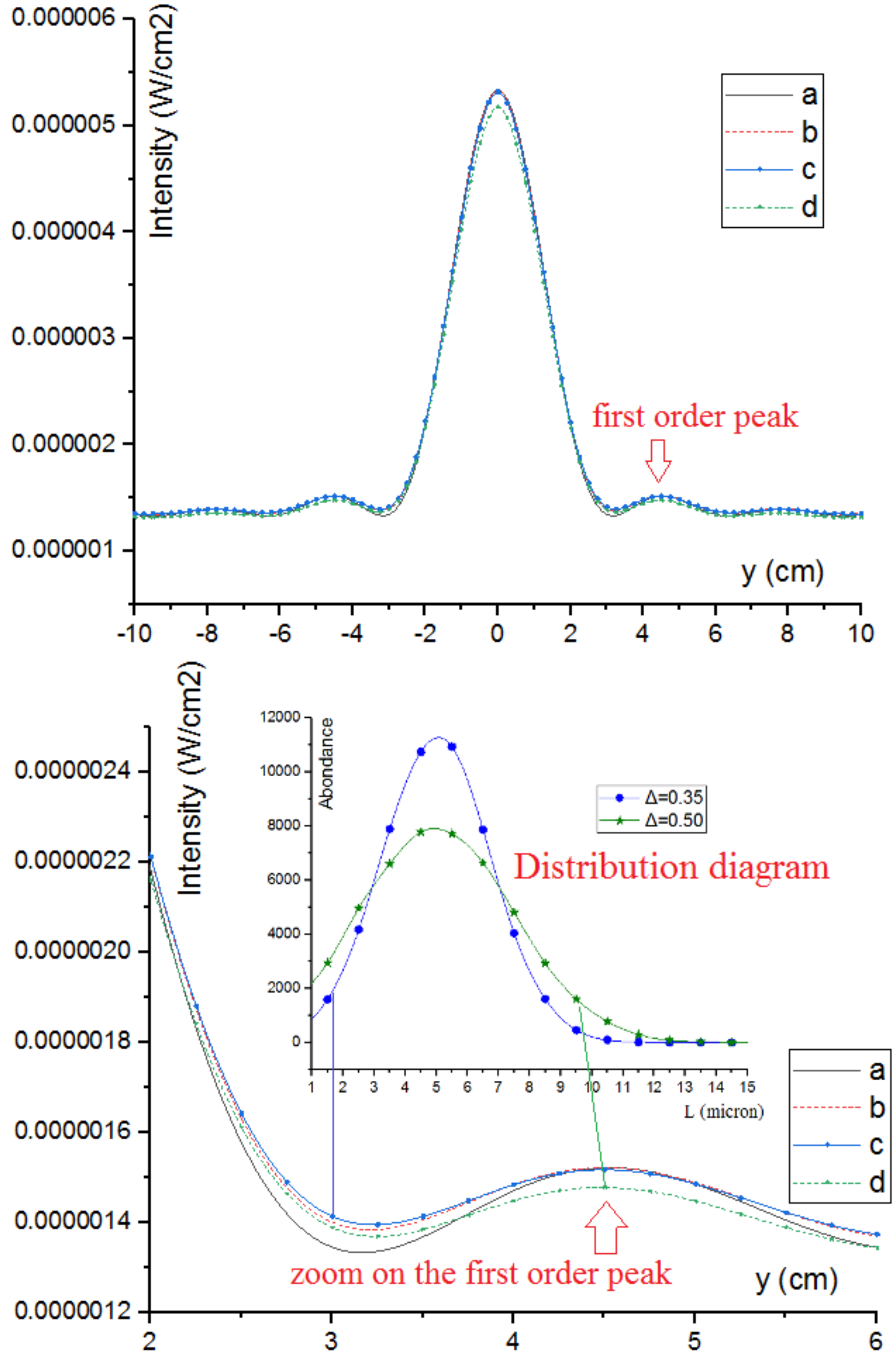

Fig.6. diffraction pattern of a single slit with $b$=50 micron illuminated by a He-Ne laser beam with wavelength of 633 nm at far-field $r$=250cm: (a). perfect coherence, (b). Partial coherence with constant $l_0 = 5$ micron ($\eta = 10$) and Δ=0, (c). Partial coherence with a Gaussian distribution $l_0 = 5$ micron ($\eta = 10$) and Δ=0.35, (d). Partial coherence with a Gaussian distribution $l_0 = 5$ micron ($\eta = 10$) and Δ=0.50. Intent shape is the statistical distribution of the coherence length.

Now we study the effects of partial temporal decoherence on the multi-slits diffraction patterns through changing the coherence length $l_0$. In Fig.(7) the diffraction pattern of two slits is plotted for different values of $l_0$ and for $\lambda = 633$ nm. For a double slits, we expect to have $2(a/b)-1$ interference peaks in the central diffraction peak [10,11]. As it is observed from Fig.(7), our model satisfies this requirement. On the other hand, for values of $\eta = b/l_0$ more than 1, the effects of partial temporal coherence begin to appear, in other words the effects of partial temporal decoherence show themselves for $\eta > 1$.

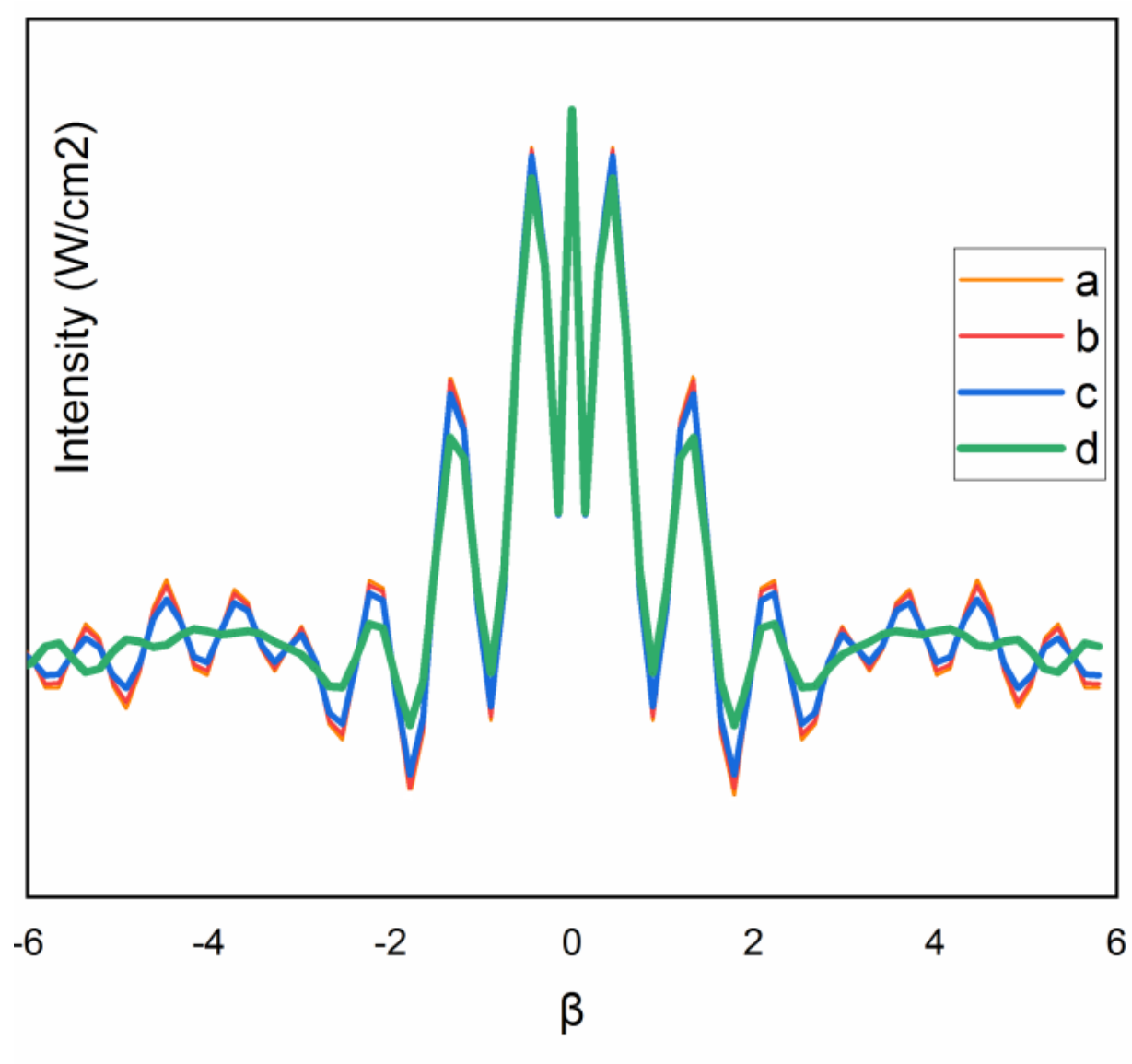

Fig.(7). Intensity distribution from a double slits with a=90 microns, b=30 microns and different numbers of coherence length: (a) perfect temporal coherence ($\eta \to 0$), (b) $\eta = 1$ (c) $\eta = 3$ and (d) $\eta = 10$.

Figs (8) and (9) show diffraction pattern of a triple slits. It is seen that the different orders of diffraction become sharper when the number of slits increases, which is also expected in common models (perfect coherence)[10,11]. Also, again for values of $\eta = b/l_0$ more than 1, the effects of partial temporal coherence begin to appear. It is an interesting result, because it shows that for multi slits, the effects of decoherence manifest themselves even for $l_0$ which is of the order of the slits dimension. Strictly speaking, the interaction of the coherence length and the slits begins when the coherence length is shorter than the size of the slits. For other values of n, the same results are obtained.

As an important note, increasing the decoherence parameter, influences the intensity distribution. In multi-slits, these effects seem to be destructive as shown in Figures (7-9). Figures show that the role of the temporal decoherence factor becomes even more dominant when the number of slits are increasing. It should be noted that destructive effects of partial coherence beam on the interference patterns have been approved and documented before, but here it is developed for more general case of multi-slits [23-24].

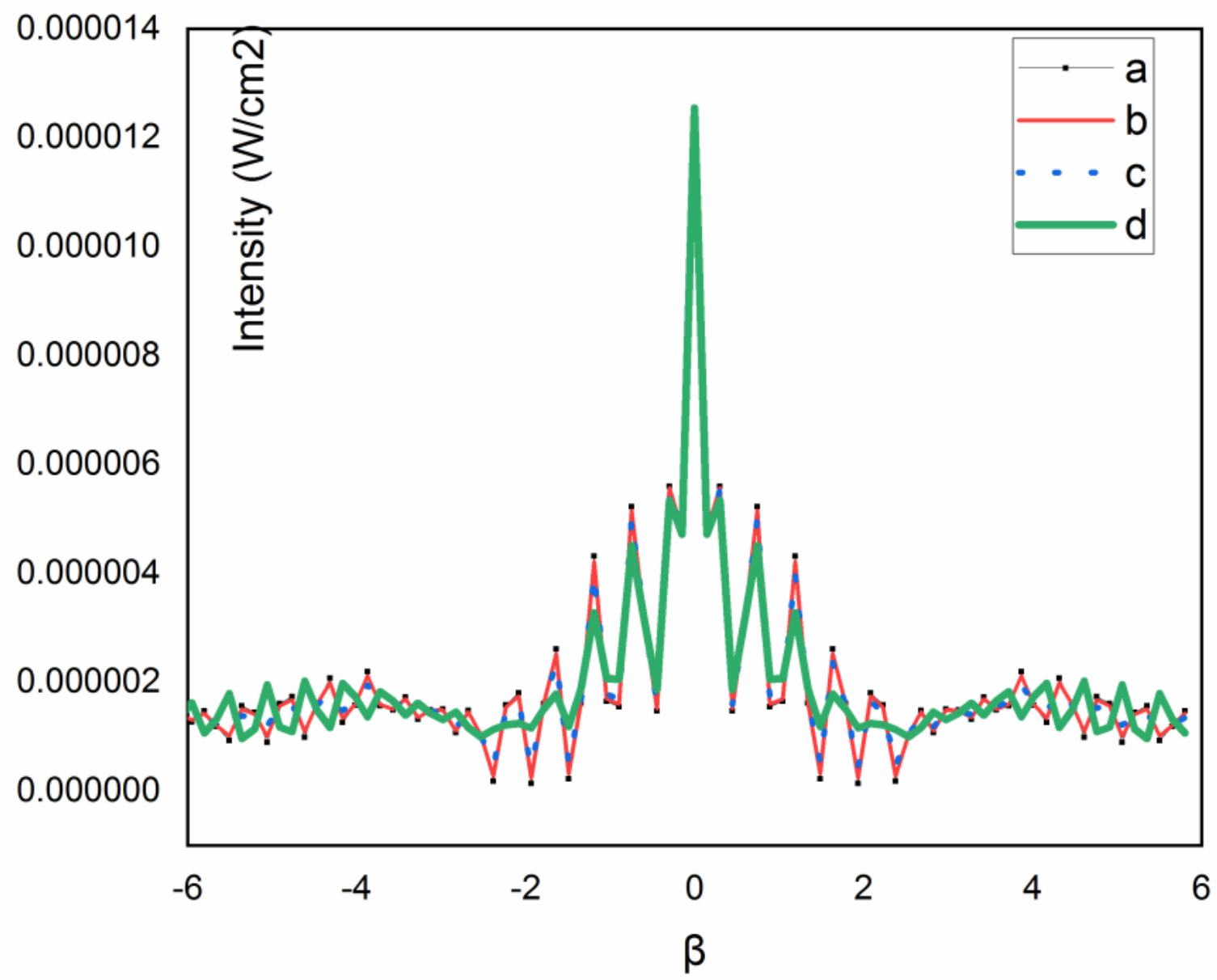

Fig. (8). Intensity distribution from triple-slits with a=90 microns, b=30 microns and different numbers of coherence length: (a) perfect temporal coherence ($\eta \to 0$), (b) $\eta = 1$ (c) $\eta = 3$ and (d) $\eta = 10$.

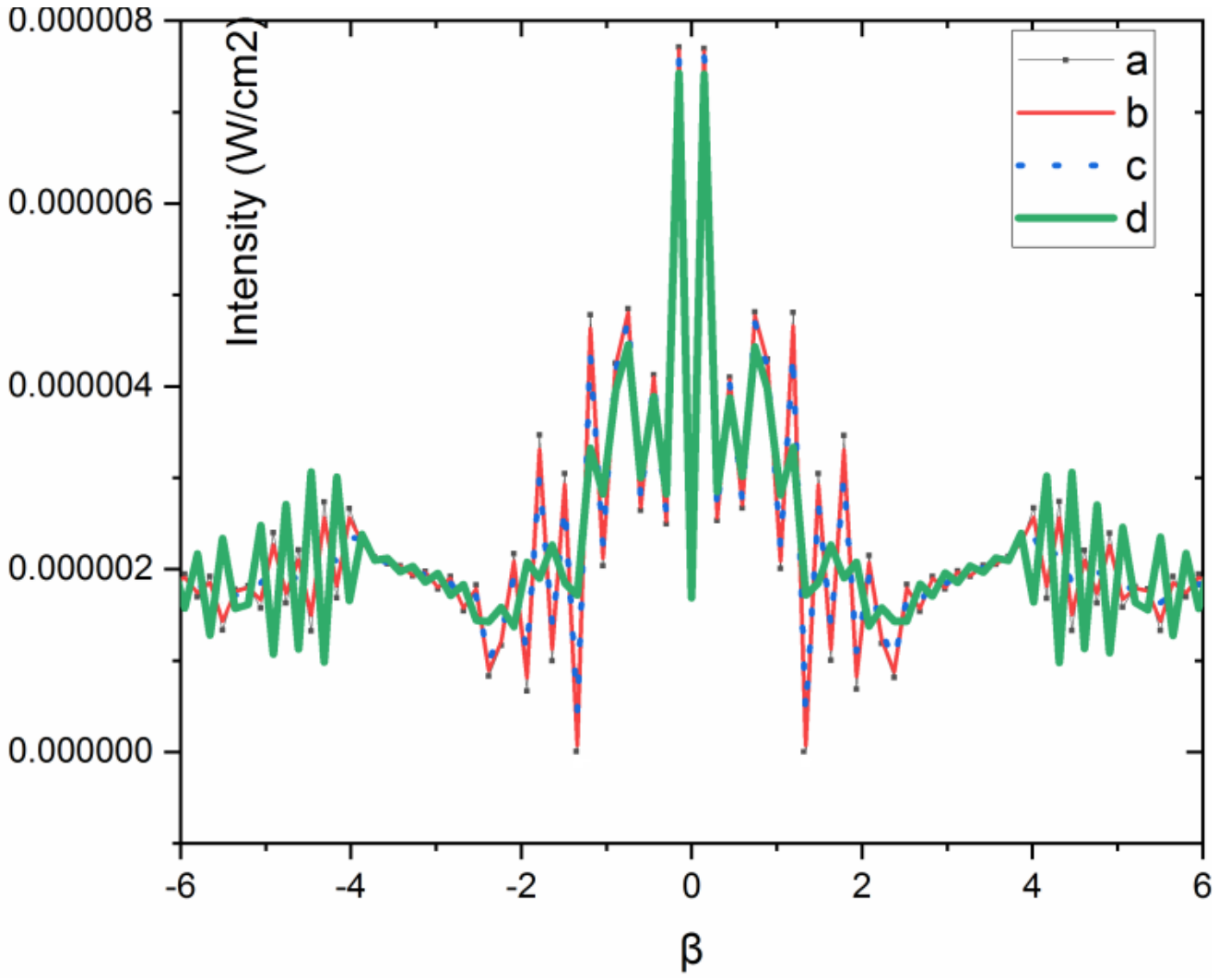

Fig.(9). Intensity distribution from four slits with a=90 microns, b=30 microns and different numbers of coherence length: (a) perfect temporal coherence ($\eta \to 0$). (b) $\eta = 1$ (c) $\eta = 3$ and (d) $\eta = 10$.

Finally, we discuss the effects of Gaussian distribution of the coherence length on the diffraction from multi-slits. For the sake of brevity, we consider only the case of triple-slits. In Fig.(10) the case (d) of Fig.(8) (constant coherence length with $\eta$=10) is compared with the case of Gaussian distribution with Δ=0.5 ( Fig.(6) d). It is worth mentioning that our results show that by controlling the width of Gaussian distribution of the coherence length one can reduce the destructive effects on the first order of diffraction pattern. This may be caused by the contributions of the part of the distribution that has longer coherence lengths and compensates the destructive effects of shorter coherence lengths. These consequences can be extended to different number of slits and values of Δ. The results may be of practical importance, especially when we want to compensate the destructive effects of short coherence length in the diffraction orders of a grating. These considerations could be useful in the design of spectroscopic instruments.

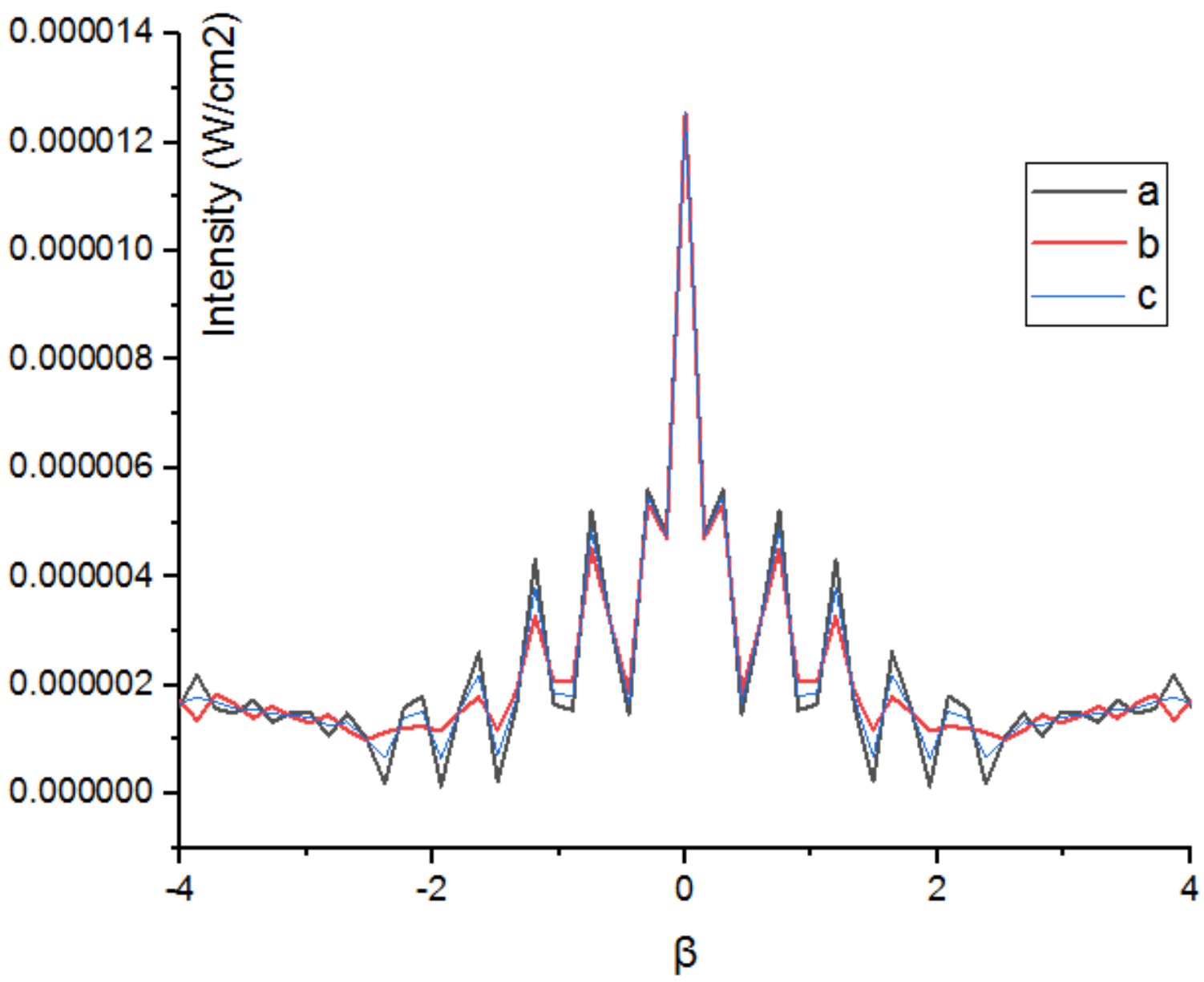

Fig. (10). Intensity distribution from triple-slits with a=90 microns, b=30 microns and different numbers of coherence length: (a) perfect temporal coherence ( $\eta \to 0$ ), (b) $\eta = 10$ and $l_0$ is constant, (c) $\eta = 10$ and $l_0$ has a Gaussian distribution

## 4. Conclusion

We have generalized the Fraunhofer diffraction from a single slit illuminated by a partial temporal coherent beam previously studied in our work [19] to the case of multi-slits, both from theory and simulation point of view. To achieve this goal, theoretical and numerical methods have been developed for the simulation of the diffraction pattern of multi-slits which is also used to study the effects of coherence length on the pattern. The diffraction pattern is investigated for different values of the decoherence parameter $\eta$. It is shown that for multi-slits, the temporal decoherence effects appear for $\eta \geq 1$. The results of our study reproduce the outcome of previous studies for perfect temporal case, when coherence length tends to infinity. The more real and important case that the coherence length has Gaussian distribution is also studied and the effects of this distribution are investigated on the diffraction pattern, using Monte Carlo method.

In conclusion our results may benefit optical engineering technologies in particular designing spectrometers. By adjusting the decoherence parameter, one can adjust the brightness of the first and second orders of diffraction peaks which may create new features in spectroscopy.

**Acknowledgement**

We also acknowledge Dr. Asghar Moulavi Nafchi, from the department of English at HSU, for proofreading of the manuscript.

**Appendix A:**

In Eq. (1), the first term is the summation of $N$ self-product of different segments at the point $P$. To find an explicit form for it, we note that the electric field of $j$'th segment with $\Delta y_j$ width at distance $r$ is given by [10]:

$$(\frac{1}{b\ r}\lim_{N'\to\infty}\varepsilon_0 N'\ \Delta y_j)\ \text{(A.1)}$$

So we have:

$$I_s = \lim_{N\to\infty}\sum_{j=1}^{N-1} I_j = \lim_{N\to\infty}\sum_{j=1}^{N-1}\left(\frac{1}{b^2r^2}\lim_{N'\to\infty}\varepsilon_0^2 N'^2 \Delta y_j^2\right) = \lim_{N\to\infty}\left(\frac{E_L}{r}\right)^2\sum_{j=1}^{N-1}(\Delta y_j^2) \quad \text{(A.2)}$$

Also we would have:

$$\sum_{j=1}^{N-1}(\Delta y_j^2) = (\Delta y_1^2 + \Delta y_2^2 + \ldots. + \Delta y_{N-1}^2) =$$
$$(\Delta y_1 + \Delta y_2 + \ldots. + \Delta y_{N-1})^2 - 2\Delta y_1 \Delta y_2 - 2\Delta y_1 \Delta y_3 - \ldots \quad \text{(A.3)}$$

as we know, $\Delta y_1 = \Delta y_2 = \ldots = \Delta y_{N-1}$. Thus:

$$\sum_{j=1}^{N-1}(\Delta y_j^2) = (\Delta y_1^2 + \Delta y_2^2 + \ldots. + \Delta y_{N-1}^2) =$$
$$(\Delta y_1 + \Delta y_2 + \ldots. + \Delta y_{N-1})^2 - 2(\Delta y_1^2 + \Delta y_2^2 + \ldots. + \Delta y_{N-2}^2) \approx \quad \text{(A.4)}$$
$$(b)^2 - 2\left(\sum_{j=1}^{N-1}(\Delta y_j^2)\right)$$

This equation gives:

$$\sum_{j=1}^{N-1}(\Delta y_j^2) = b^2/3$$

So, we arrive at the following expression:

$$I_s = \left(\frac{E_L}{r}\right)^2 \frac{b^2}{3} \quad \text{(A.5)}$$

Which is the first term in Eq.(2).

**Appendix B:**

The analytical solution of the integral in Eq.(2), is as follows:

$$\int (1-\alpha y)(1-\beta y)\cos(\gamma y)dy =$$

$$(\frac{1}{\gamma}-\frac{2\alpha\beta}{\gamma^3})\sin(\gamma y)-(\frac{\alpha+\beta}{\gamma})y\sin(\gamma y)-(\frac{\alpha+\beta}{\gamma^2})\cos(\gamma y)+ \qquad \text{(B.1)}$$

$$(\frac{\alpha\beta}{\gamma})y^2\sin(\gamma y)+(\frac{2\alpha\beta}{\gamma^2})y\cos(\gamma y)$$